\documentclass{moriond}

\def\be{\begin{equation}}
\def\ee{\end{equation}}
\def\bea{\begin{eqnarray}}
\def\eea{\end{eqnarray}}

\usepackage{amsmath,amssymb}
\usepackage{bm}
\usepackage{graphicx} 
\usepackage{hyperref}
\usepackage{mathrsfs}
\usepackage{multirow}    
\usepackage{dsfont}    
\usepackage{bbm}

\newcommand{\Id}{\mathbbm 1}

\usepackage{ulem}
\usepackage[usenames]{color}

\allowdisplaybreaks
\newcommand{\Photo}{}
\newcommand{\ud}{\mathrm{d}}

\begin{document}
\vspace*{4cm} \title{BIMETRIC GRAVITY AND DARK MATTER}

\author{Laura BERNARD and Luc BLANCHET}

\address{$\mathcal{G}\mathbb{R}\varepsilon{\mathbb{C}}\mathcal{O}$
  Institut d'Astrophysique de Paris, UMR 7095, CNRS,\\ Sorbonne
  Universit{\'e}s \& UPMC Univ Paris 6, 98\textsuperscript{bis}
  boulevard Arago, 75014 Paris, France}

\author{Lavinia HEISENBERG}

\address{Nordita, KTH Royal Institute of Technology and Stockholm
  University, \\Roslagstullsbacken 23, 10691 Stockholm, Sweden, and
  \\Department of Physics \& The Oskar Klein Centre, \\AlbaNova
  University Centre, 10691 Stockholm, Sweden}

\maketitle

\abstracts{We review some recent proposals for relativistic models of
  dark matter in the context of bimetric gravity. The aim is to solve
  the problems of cold dark matter (CDM) at galactic scales, and to
  reproduce the phenomenology of the modified Newtonian dynamics
  (MOND), while still being in agreement with the standard
  cosmological model $\Lambda$-CDM at large scales. In this context a
  promising alternative is dipolar dark matter (DDM) in which two
  different species of dark matter particles are separately coupled to
  the two metrics of bigravity and are linked together by an internal
  vector field. The phenomenology of MOND then results from a
  mechanism of gravitational polarization. Probably the best
  formulation of the model is within the framework of recently
  developed massive bigravity theories. Then the gravitational sector
  of the model is safe by construction, but a ghostly degree of
  freedom in the decoupling limit is still present in the dark matter
  sector. Future work should analyze the cosmological solutions of the
  model and check the post-Newtonian parameters in the solar system.}

\section{Introduction and motivation}
\label{sec:intro}

\subsection{Problem of dark matter at galactic scales}

The standard model of cosmology $\Lambda$-CDM is widely held to be an
excellent description of reality at large cosmological
scales. Impressive observational successes of this model include the
fit of the anisotropies of the cosmic microwave background (CMB), the
baryon acoustic oscillations, the formation of large scale structures
and the accelerated expansion of the Universe. However some
fundamental issues remain: (i) The measured value of the cosmological
constant $\Lambda$ looks unnatural from a quantum field perspective;
(ii) The weakly interacting particles envisaged as candidates for the
cold dark matter (CDM) are still undetected in the laboratory; (iii)
The model $\Lambda$-CDM falls short in explaining the observed
regularities in the properties of dark matter halos around galaxies.

Regarding point (iii) we have in mind the baryonic Tully-Fisher
relation between the observed luminous mass and the asymptotic
rotation velocity of spiral galaxies~\cite{McG11}, the analogous
Faber-Jackson relation for elliptical galaxies~\cite{Sand10}, and,
very important, the tight correlation between the mass discrepancy
(i.e. the presence of dark matter) and the involved acceleration
scale~\cite{FamMcG12}. In the prevailing view~\cite{SilkM12}, these
issues should be resolved once we understand the complicated baryonic
processes (e.g. supernova winds and outflows from a central
supermassive black hole) that affect galaxy formation and
evolution. Note that with such foreseen explanation, conclusive tests
of the $\Lambda$-CDM model become difficult since any new failure can
be attributed to some further unknown aspect of baryonic physics. More
importantly, this explanation is challenged by the fact that galactic
data are --- mysteriously enough --- in excellent agreement with the
MOND (MOdified Newtonian Dynamics) empirical non relativistic
formula~\cite{Milg1,Milg2,Milg3}.

A relativistic MOND theory is required to address issues concerning
cosmology and gravitational lensing. Many relativistic extensions for
MOND have been proposed, including a tensor-vector-scalar (TeVeS)
theory~\cite{Bek04,Sand05}, a bimetric theory~\cite{bimond1},
non-canonical Einstein-{\AE}ther theories~\cite{ZFS07,Halle08}, a
Galileon theory~\cite{BDgef11}, a Khronon theory~\cite{BM11}, a
modified dark matter theory~\cite{BL08,BL09,BLLM13}. Most theories
have difficulties at reproducing the cosmological observations,
notably the full spectrum of CMB anisotropies. An exception is the
modified dark matter approach~\cite{BL08,BL09,BLLM13} which agrees
with the model $\Lambda$-CDM at first order cosmological
perturbations. This approach, also called dipolar dark matter (DDM),
is motivated by the dielectric analogy of MOND --- that MOND
represents the gravitational analogue of the Gauss law of
electrostatics in dielectric non-linear
materials~\cite{B07mond,BBwag}. A natural formulation of the model is
based on a bimetric extension of general relativity (GR)~\cite{BB14},
where two species of dark matter particles are respectively coupled to
the two metrics, and are linked by an internal vector field generated
by the mass of these particles. In this model the phenomenology of
MOND emerges naturally and elegantly from a mechanism of gravitational
polarization.

\subsection{Bimetric massive gravity theories}
\label{sec:review}

Bimetric theories have been extensively investigated in the quest of a
consistent massive gravity theory. From a theoretical point of view,
the existence of a graviton mass is a very important fundamental
question. At the linear level, there is a unique mass term which
ensures the absence of ghosts in the theory, namely the Fierz-Pauli
action~\cite{Fierz:1939ix}. Even though theoretically consistent, this
linear theory suffers from the van Dam-Veltman-Zakharov (vDVZ)
discontinuity~\cite{vanDam:1970vg,Zakharov:1970cc}, which reflects the
fact that one does not recover GR in the limit of vanishing graviton
mass. In other words, this discontinuity can be traced to the coupling
of the additional longitudinal graviton to the matter field in the
mass going to zero limit. Vainshtein very soon realized that the
nonlinearities of the theory become actually stronger in the vanishing
mass limit~\cite{Vainshtein:1972sx}, pointing that these
nonlinearities might cure the vDVZ discontinuity, which required the
non-linear completion of the Fierz-Pauli theory. Unfortunately, this
task seemed to face the inevitable problem of reintroducing the
ghost-like instability~\cite{Boulware:1973my}.

Recent joint effort to construct a consistent ghost-free non-linear
theory for massive gravity gave fruitful
results~\cite{deRham:2010ik,deRham:2010kj,Hassan:2011vm,Hassan:2011hr},
which have initiated a revival of interests. The theory can be further
generalized to bigravity~\cite{Hassan:2011zd}, or
multigravity~\cite{Hinter12}, by the inclusion of the corresponding
additional kinetic terms. Another fundamental question in the context
of massive gravity is the consistent coupling to the matter fields.
If one couples the matter fields in a naive way additively to both
metrics simultaneously, this immediately reintroduces the
Boulware-Deser (BD) ghost~\cite{dRHRa,Yamashita:2014fga}. Moreover,
quantum corrections at one loop dictate that the only consistent way
of coupling the matter sector to the two metrics is either through a
minimal coupling to just one metric (which preserves the technical naturalness of the
theory~\cite{deRham:2012ew,deRham:2013qqa,Heisenberg:2014rka}), or
through a composite effective metric built out of both metrics in a
very specific way~\cite{dRHRa,Heisenberg:2014rka,dRHRb}. An important
consequence of the coupling through the effective metric is a possible
way out of the no-go result for the flat
Friedmann--Lema\^itre--Robertson--Walker (FLRW) metric together with
the propagation of the five physical degrees of freedom of the
graviton sector without introducing ghost and gradient
instabilities~\cite{Gumrukcuoglu:2014xba}.

Massive gravity is replete of phenomenology. Especially, its potential
application in cosmology received much attention. Even if the
decoupling limit of the theory admits self-accelerating
solutions~\cite{deRham:2010tw}, the full theory with Minkowski
reference metric suffers from the no-go result for flat FLRW
solution~\cite{PhysRevD.84.124046}. The cosmology of the bigravity
theory has more freedom and features due to the dynamics of the
reference
metric~\cite{Volkov:2011an,vonStrauss:2011mq,Comelli:2011zm}. Assuming
that the matter fields couple minimally to one metric and further
assuming that the mass of the graviton is of the same order as the
Hubble parameter today, the theory admits several interesting branches
of solutions.

\subsection{Content and summary}

In Sec.~\ref{sec:bimetric} of this paper we shall review the initial
bimetric model~\cite{BB14} and the broad range of phenomenology it is
able to predict. On the bad side of this model, we shall also discuss,
in Sec.~\ref{sec:mini}, with the help notably of the minisuperspace of
the model, the likely presence of ghost instabilities. This will
motivate the redefinition, in Sec.~\ref{sec:massivegrav}, of the
gravitational sector of the model~\cite{BB14} in a way to make it
consistent with the beautiful framework of massive bigravity
theories~\cite{BH15a,BH15b}. Gladly, this move will substantially
simplify the model. On the other hand, the dark matter sector will
essentially remain the same as in~\cite{BB14}. The mechanism of
gravitational polarization, and recovery of the MOND equation, checked
in Sec.~\ref{sec:polar}, thus appear as a natural consequence of
massive bigravity theory for this type of matter. Concerning ghosts,
the new model is safe in the gravitational sector (by construction),
but a ghostly degree of freedom in the decoupling limit is
reintroduced in the dark matter sector~\cite{BH15b}. A crucial
question to address in future work is whether the polarization
mechanism can be realized in absence of ghosts.

\section{Bimetric theory with two dark matter species}
\label{sec:bimetric}

\subsection{Relativistic action}
\label{sec:action}

 A relativistic model involving, in addition to the ordinary matter
 simply described by baryons, two species of dark matter particles,
 was proposed in~\cite{BB14}. A vector field $\mathcal{A}_\mu$ is
 sourced by the mass currents of dark matter and dubbed
 \textit{graviphoton}. This vector field is crucial in order to ensure
 the stability of the dipolar medium. The gravitational sector is
 composed of two dynamical Lorentzian metrics $g_{\mu\nu}$ and
 $f_{\mu\nu}$. The baryons (representing in fact the full standard
 model of particle physics) are coupled in the usual way to the metric
 $g_{\mu\nu}$. The two species of dark matter particles are
 respectively coupled to the two metrics $g_{\mu\nu}$ and
 $f_{\mu\nu}$. The gravitational-plus-matter action of the
 model~\cite{BB14} reads (in geometrical units $G=c=1$)
\begin{align}\label{action} S &= \int\mathrm{d}^{4}x\left\{
    \sqrt{-g}\left(\frac{R_g-2\lambda_g}{32\pi}-\rho_\text{bar}-\rho_g\right)
    + \sqrt{-f}\left(\frac{R_f-2\lambda_f}{32\pi} -
      \rho_f\right)\right. \nonumber\\ &
  \left.\qquad\qquad+\sqrt{-\mathcal{G}_\text{eff}}
    \left[\frac{\mathcal{R}_\text{eff}-2\lambda_\text{eff}}{16\pi
        \varepsilon}
      +\bigl(\mathcal{J}_g^\mu-\mathcal{J}_f^{\mu}\bigr)\mathcal{A}_\mu
      + \frac{a_{0}^{2}}{8\pi}\,\mathcal{W}(\mathcal{X})\right]
  \right\} \,.\end{align}
Here $R_g$, $R_f$ and $\mathcal{R}_\text{eff}$ are the Ricci scalars
of the $g$ and $f$ metrics, and of an effective composite metric
$\mathcal{G}^\text{eff}_{\mu\nu}$ defined non-perturbatively from
$g_{\mu\nu}$ and $f_{\mu\nu}$ by\,\footnote{Actually the effective
  composite metric was defined in~\cite{BB14} by the different
  condition
$$\mathcal{G}^\text{eff}_{\mu\nu} =
  \mathcal{G}_\text{eff}^{\rho\sigma} g_{\rho\mu}\,f_{\nu\sigma} =
  \mathcal{G}_\text{eff}^{\rho\sigma} g_{\rho\nu}\,f_{\mu\sigma}\,.$$
  It has been shown~\cite{BH15a} that this condition is equivalent to
  the requirement~\eqref{Geff}.}
\begin{equation}\label{Geff}
\mathcal{G}^\text{eff}_{\mu\nu} =
g_{\mu\rho}X^\rho_\nu=f_{\mu\rho}Y^\rho_\nu\,,
\end{equation}
where the square root matrix is defined by $X=\sqrt{g^{-1}f}$,
together with its inverse $Y=\sqrt{f^{-1}g}$. Notice that
$\mathcal{G}^\text{eff}_{\mu\nu}$ can be shown to be automatically
symmetric~\cite{Baccetti:2012re,Hassan:2012wr}. The dimensionless
coupling constant $\varepsilon$ measures the strength of the
interaction between the two sectors $g$ and $f$. We inserted three
different cosmological constants $\lambda_g$, $\lambda_f$ and
$\lambda_\text{eff}$ in the $g$, $f$ and interaction sectors. They
will be ultimately related to the observed cosmological constant
$\Lambda$.

Baryons and dark matter particles are described by pressureless fluids
with conserved scalar densities $\rho_\text{bar}$, $\rho_g$ and
$\rho_f$. In addition $\mathcal{J}_g^{\mu}$ and $\mathcal{J}_f^{\mu}$
stand for the mass currents of the two types of dark matter particles,
defined by
\begin{equation}
\sqrt{-\mathcal{G}_\text{eff}}\mathcal{J}_g^{\mu}=\sqrt{-g}
J_g^{\mu}\,,\qquad\sqrt{-\mathcal{G}_\text{eff}}\mathcal{J}_f^{\mu}=\sqrt{-f}
J_f^{\mu}\,,
\end{equation}
where $J_g^{\mu}=\rho_g u^\mu_g$ and $J_f^{\mu}=\rho_f u^\mu_f$ are
the conserved dark matter currents associated with the respective
metrics $g$ and $f$, thus obeying $\nabla^g_{\mu}J_g^{\mu}=0$ and
$\nabla^f_{\mu}J_f^{\mu}=0$.

The vector field $\mathcal{A}_\mu$ obeys a non-canonical kinetic term
$\mathcal{W}(\mathcal{X})$ where
\begin{equation}\label{calX} 
\mathcal{X} = -\frac{\mathcal{G}_\text{eff}^{\mu\rho}
  \mathcal{G}_\text{eff}^{\nu\sigma}\mathcal{F}_{\mu\nu}
  \mathcal{F}_{\rho\sigma}}{2a_{0}^{2}}\,,
\end{equation}
with $\mathcal{F}_{\mu\nu}=\partial_\mu \mathcal{A}_\nu-\partial_\nu
\mathcal{A}_\mu$. Note that the vector field strength in~\eqref{calX}
has been rescaled by the MOND acceleration $a_0\simeq
1.2\,10^{-10}\,\text{m}/\text{s}^2$. The function
$\mathcal{W}(\mathcal{X})$ is determined phenomenologically, but in
principle it should be interpreted within some more fundamental
theory. In the limit $\mathcal{X}\ll 1$, which corresponds to the MOND
weak-acceleration regime below $a_0$, we impose
\begin{equation}\label{W0} 
\mathcal{W}(\mathcal{X})= \mathcal{X}-\frac{2}{3} \mathcal{X}^{3/2} +
\mathcal{O}\left(\mathcal{X}^2\right)\,.
\end{equation}
On the other hand we also impose that when $\mathcal{X}\gg 1$,
corresponding to the strong-acceleration regime much above $a_0$,
\begin{equation}\label{Winf} 
\mathcal{W}(\mathcal{X})= A + \frac{B}{\mathcal{X}^{b}} +
o\!\left(\frac{1}{\mathcal{X}^{b}}\right)\,,
\end{equation}
where $A$ and $B$ are constants and $b>0$.

\subsection{Phenomenological predictions}
\label{sec:phenom}

We now review the rich phenomenology of this model, which is quite
successful at different scales and in different regimes. We refer
to~\cite{BB14} for the details.

\begin{itemize}
\item \textit{Cosmology.} We study the cosmology of the model by
  expanding the two metrics around two homogeneous and isotropic FLRW
  background metrics,
\begin{subequations}\begin{eqnarray}\label{FLRW}
\mathrm{d}\bar{s}_g^{2} &=& a_g^2\left[-\mathrm{d}\eta^2 +
  \gamma_{ij}\,\mathrm{d}x^{i}\,\ud x^{j}\right]\,,
\\ \mathrm{d}\bar{s}_f^{2} &=& a_f^2\left[-\mathrm{d}\eta^2 +
  \gamma_{ij}\,\mathrm{d}x^{i}\,\ud x^{j}\right]\,,
\end{eqnarray}
\end{subequations}
where $\eta$ is the conformal time and $a_g$ and $a_f$ are the two
different scale factors. We show that the consistency of the equations
in the background, where the two metrics superpose, is ensured
provided that the background densities obey,
\begin{equation}
\bar{\rho}_b=\frac{(\alpha-1)(\varepsilon-1)}{\alpha+\varepsilon}\bar{\rho}
\,,
\end{equation}
where $\alpha=a_g/a_f$ is constant, and $\bar{\rho}$ is the common
density of the two dark matter fluids in the background. The
cosmological constants should also be related to the observed
cosmological constant $\Lambda$, through the relations,
\begin{equation}
\lambda_g=\Lambda\,,\qquad \lambda_f=\alpha^2\Lambda\,, \qquad
\lambda_\text{eff}=\alpha\Lambda \,.
\end{equation}
Then, at first order cosmological perturbations around the FLRW
metrics, we define the $g$-sector as being the observable one because
it is where the baryons live and light propagates, and all
observations take place. We thus study the perturbation equations in
this sector. We define some effective dark matter quantities that
modify the $g$-type dark matter particles taking into account their
interaction with the other sector. Using these new variables we find
that the perturbation equations in the $g$ sector take exactly the
same form as those of the $\Lambda$-CDM model. Thus the model turns
out to be indistinguishable from $\Lambda$-CDM up to first order
perturbation, and is then fully consistent with the observed
fluctuations of the CMB.
\item \textit{MOND.} At galactic scales, in a regime of small
  accelerations $a\ll a_0$, the potential function $\mathcal{W}$ takes
  the form~\eqref{W0}. Based on a particular solution for the equation
  of the vector field $\mathcal{A}_\mu$, the two dark matter fluids
  can be described as a polarizable dipolar medium. The MOND equation
  is then recovered as a result of a mechanism of gravitational
  polarization which appears as a natural consequence of the
  model. Moreover the dipolar dark matter medium undergoes stable
  plasma-like oscillations. We shall give some more details of this
  polarization mechanism in Sec.~\ref{sec:polar} at the occasion of
  the next model based on massive bigravity. However note that the
  next model has essentially the same predictions with regards to MOND
  phenomenology as the model~\cite{BB14}. There is though a
  difference, in that in the model~\cite{BB14} we have to assume that
  the coupling constant $\varepsilon$ in the action~\eqref{action}
  tends to zero, while no particular requirement will be necessary in
  the next model.
\item \textit{Solar System.} As we modify GR we have to check the
  post-Newtonian limit in the Solar System, for which the potential
  function $\mathcal{W}$ in the action takes the form~\eqref{Winf}. We
  have shown that when we expand the model at first post-Newtonian
  order, the parametrized post-Newtonian (PPN) parameters are exactly
  the same as in GR. A crucial point for this test is a non-linear
  effect present in our definition of the effective
  metric~\eqref{Geff} and which permits to recover the correct value
  for the parameter measuring the amount of non-linearity,
  $\beta^\text{PPN}=1$. Thus the model passes the Solar System tests
  and is viable.
\end{itemize}

\subsection{Minisuperspace and ghosts}
\label{sec:mini}

Unfortunately, the previous model proposed in~\cite{BB14} suffers from
an harmful ghost instability. The source of its origin is
multifaceted. First of all, the presence of the square root of the
determinant of $\mathcal{G}_\text{eff}$ in the action~\eqref{action}
corresponds to ghostly potential interactions,
\begin{equation}
\sqrt{-\mathcal{G}_\text{eff}} = \sqrt{\sqrt{-g}\sqrt{-f}}\,.
\end{equation}
Already at the linear order around a flat background, posing
$g_{\mu\nu}=(\eta_{\mu\nu}+h_{\mu\nu})^2$ and
$f_{\mu\nu}=(\eta_{\mu\nu}+\ell_{\mu\nu})^2$, these potential
interactions do not preserve the Fierz-Pauli tuning, with $[\cdots]$
denoting the trace as usual,
\begin{align}
\sqrt{-\mathcal{G}_\text{eff}} &= 1+\frac{1}{2}\bigl[h + \ell\bigr] +
\frac{1}{8}\Bigl(\bigl[h+\ell\bigr]^2
-2\bigl[h^2+\ell^2\bigr]\Bigr)\nonumber\\ &+
\!\!\frac{1}{48}\Bigl(\bigl[h+\ell\bigr]^3
-6\bigl[h+\ell\bigr]\bigl[h^2+\ell^2]+8\bigl[h^3+\ell^3\bigr]\Bigr) +
\cdots \,,
\end{align}
yielding a linear ghost instability already present at the scale
\begin{equation}
  m^2M_{\text{Pl}}^2\sqrt{-\mathcal{G}_\text{eff}}\sim \frac{m^2
    M_{\text{Pl}}^2(\Box\pi)^2}{\Lambda_3^6}=
  \frac{(\Box\pi)^2}{m^2}\,,
\end{equation}
where $\Lambda_3=(M_{\text{Pl}}m^2)^{1/3}$ and $\pi$ encodes the
helicity-0 degree of freedom of the massive graviton with mass
$m$. The ghost corresponds to a very light degree of freedom, and
hence the theory cannot be used as an effective field theory. The
other source of ghostly interactions in the model~\cite{BB14} is
originated in the presence of three kinetic terms. Studying the theory
in the mini-superspace immediately reveals the ghostly interactions of
the considered kinetic terms. The dangerous sub-Lagrangian due to
kinetic interactions is given by
\begin{equation}
\mathcal{L}^\text{eff}_\text{kin} = \frac{M_{\text{Pl}}^2}{2}
\sqrt{-g} R_g + \frac{M_\text{eff}^2}{2}\sqrt{-\mathcal{G}_\text{eff}}
\mathcal{R}_\text{eff}\,.
\end{equation}
As a first diagnostic, we can assume the two metrics to be of the
mini-superspace form
\begin{subequations}
\begin{eqnarray}
\ud s_g^2&=&g_{\mu\nu} \ud x^\mu \ud x^\nu = -n_g^2 \ud t^2 + a_g^2
\ud x^2\,, \\ \ud s_f^2&=&f_{\mu\nu} \ud x^\mu \ud x^\nu = -n_f^2 \ud
t^2 + a_f^2 \ud x^2\,,
\end{eqnarray}
\end{subequations}
where $n_g$, $n_f$ are the lapse functions and $a_g$, $a_f$ are the
scale factors of the respective metrics. The effective metric
$\mathcal{G}_\text{eff}$ in the mini-superspace becomes
\begin{eqnarray}
\ud s_{\mathcal{G}_\text{eff}}^2&=&\mathcal{G}^\text{eff}_{\mu\nu} \ud
x^\mu \ud x^\nu = -n_\text{eff}^2 \ud t^2 + a_\text{eff}^2 \ud x^2\,,
\end{eqnarray}
where the effective lapse and effective scale factor correspond to
$n_\text{eff}=\sqrt{n_gn_f}$ and $a_\text{eff}=\sqrt{a_ga_f}$. We
compute the conjugate momenta for the scale factors and get
\begin{align}
p_g &=-6M_{\text{Pl}}^2a_g^2H_g-\frac{3}{2}M_\text{eff}^2a_f
\frac{a_\text{eff}}{n_\text{eff}}\bigl(H_gn_g+H_fn_f\bigr)\,,
\nonumber\\ p_f &=-\frac{3}{2}M_\text{eff}^2a_g
\frac{a_\text{eff}}{n_\text{eff}}\bigl(H_gn_g+H_fn_f\bigr)\,,
\end{align}
with the conformal Hubble factors $H_g=\frac{\dot{a}_g}{a_gn_g}$ and
$H_f=\frac{\dot{a}_f}{a_fn_f}$ respectively. After performing the
Legendre transformation we obtain the Hamiltonian in the
mini-superspace:
\begin{align}\label{hamiltonian}
& \mathcal{H}^\text{eff}_\text{kin} =- \frac{(p_g a_g - p_f
    a_f)^2n_g}{12M_{\text{Pl}}^2a_g^3}- \frac{p_f^2a_g
    a_{\text{eff}}n_{\text{eff}}}{3 M_{\text{eff}}^2a_g^3}\,.
\end{align}
As one can see immediately, the Hamiltonian is highly non-linear in
the lapses $n_g$ and $n_f$, signalling the presence of the BD ghost
degree of freedom already in the mini-superspace. Summarising, the
model proposed in~\cite{BB14} cannot represent a consistent theory for
dipolar dark matter in bigravity due to the ghostly contribution in
form of the cosmological constant for $\mathcal{G}_\text{eff}$, and
the kinetic term $\sqrt{-\mathcal{G}_\text{eff}}
\mathcal{R}_\text{eff}$.

\section{Model based on massive bigravity theory}
\label{sec:massivegrav}

\subsection{Covariant theory}
\label{sec:cov}

The previous model~\cite{BB14} is plagued by harmful ghosts, but it
remains that the phenomenology, especially at galactic scales
(i.e. MOND), calls for a more fundamental theory. We now look for a
consistent coupling of the dark matter particles within the framework
of massive bigravity theory and the restrictions made
in~\cite{dRHRa,Heisenberg:2014rka,dRHRb} concerning matter fields. We
thus propose a new model, whose dark matter sector is essentially the
same as in the previous model~\cite{BB14}, but whose gravitational
sector is now based on ghost-free massive bigravity theory. As
bigravity theory represents essentially a unique consistent
deformation of GR, we think that the new model could represent an
important step toward a more fundamental theory of dark matter
\textit{\`a la} MOND in galactic scales.

The model is based on the bigravity-plus-matter action~\cite{BH15a,BH15b}
\begin{align}
S &= \int\mathrm{d}^{4}x\biggl\{ \sqrt{-g}\biggl(\frac{M_g^2}{2}R_g -
\rho_\text{bar}-\rho_g\biggr)
+\sqrt{-f}\biggl(\frac{M_f^2}{2}R_f-\rho_f\biggr)
\nonumber\\ &\qquad\qquad +\sqrt{-g_\text{eff}} \biggl[ \frac{m^2}{4\pi} +
  \mathcal{A}_\mu \left(j_g^\mu - \frac{\alpha}{\beta}j_f^\mu\right) +
  \frac{a_0^2}{8\pi}\,\mathcal{W}\bigl(X\bigr)
  \biggr]\biggr\}\,,\label{lagrangian}
\end{align}
where $M_g$ and $M_f$ are two coupling constants, and $m$ is the mass
of the graviton. The ghost-free potential interactions between the two
metrics $g$ and $f$ take the particular form of the square root of the
determinant of the effective metric~\cite{dRHRa,Noller:2014sta},
\begin{equation}\label{effmetric}
g^\text{eff}_{\mu\nu}=\alpha^2 g_{\mu\nu} +2\alpha\beta
\,\mathcal{G}^\text{eff}_{\mu\nu} +\beta^2 f_{\mu\nu}\,,
\end{equation}
where $\alpha$ and $\beta$ are arbitrary constants, which can always
be restricted in the model~\eqref{lagrangian} to satisfy
$\alpha+\beta=1$, and $\mathcal{G}^\text{eff}_{\mu\nu}$ denotes the
effective metric of the previous model as defined by~\eqref{Geff} in
terms of $X=\sqrt{g^{-1}f}$ and $Y=\sqrt{f^{-1}g}$. The square root of
the determinant of this effective metric admits a closed-form
expression,
\begin{equation}\label{detgeff}
\sqrt{-g_\text{eff}}=\sqrt{-g} \,\det\bigl(\alpha\Id +\beta
X\bigr)=\sqrt{-f} \,\det\bigl(\beta\Id +\alpha Y\bigr)\,,
\end{equation}
and can also be written with the help of the usual elementary
symmetric polynomials $e_n(X)$ and $e_n(Y)$ of the square root
matrices $X$ or $Y$ as
\begin{equation}\label{detgeff2}
\sqrt{-g_\text{eff}} = \sqrt{-g} \sum_{n=0}^4\alpha^{4-n}\beta^{n}
e_n(X) = \sqrt{-f} \sum_{n=0}^4\alpha^{n}\beta^{4-n} e_n(Y)\,.
\end{equation}
We see that~\eqref{detgeff}--\eqref{detgeff2} corresponds to the right
form of the acceptable potential interactions between the metrics $g$
and $f$. To recover the usual Newtonian limit for the motion of
baryons with respect to the ordinary metric $g$ we must impose
\begin{equation}\label{Nlimit} 
M_g^2 + \frac{\alpha^2}{\beta^2} M_f^2 = \frac{1}{8\pi}\,.
\end{equation}
Finally the vector field $\mathcal{A}_\mu$ is now coupled to the
metric $g^\text{eff}_{\mu\nu}$ rather than to
$\mathcal{G}^\text{eff}_{\mu\nu}$, thus
\begin{equation}\label{X} 
X = -\frac{g_\text{eff}^{\mu\rho}
  g_\text{eff}^{\nu\sigma}\mathcal{F}_{\mu\nu}
  \mathcal{F}_{\rho\sigma}}{2a_{0}^{2}}\,.
\end{equation}
However the functional form of the non-canonical kinetic term
$\mathcal{W}$ in the MOND regime is the same as in the previous model,
see~\eqref{W0}, and we could also impose~\eqref{Winf}. Finally the
mass currents to which is coupled the vector field
in~\eqref{lagrangian} are defined by
\begin{equation}\label{currents}
\sqrt{-g_\text{eff}}\,j_g^{\mu}=\sqrt{-g}
J_g^{\mu}\,,\qquad\sqrt{-g_\text{eff}}\,j_f^{\mu}=\sqrt{-f}
J_f^{\mu}\,,
\end{equation}
where, as before, $J_g^{\mu}=\rho_g u^\mu_g$ and $J_f^{\mu}=\rho_f
u^\mu_f$. However, notice the presence of the factor $\alpha/\beta$ in
the interaction term between $\mathcal{A}_\mu$ and the current
$j_f^{\mu}$ in~\eqref{lagrangian}. This factor can be interpreted as a
ratio between the gravitational charge and the inertial mass of the
$f$ particles when measured with respect to the $g$ metric.

\subsection{Gravitational polarization \& MOND}
\label{sec:polar}

We now discuss the main point of this model, which is to allow a
mechanism of gravitational polarization that permits to recover in a
straightforward way the phenomenology of dark matter at galactic
scales (MOND). In this respect the predictions of the new model are
essentially the same as those of the previous model~\cite{BB14}. 

For this purpose we are mainly interested to that particular
combination of the two metrics $g$ and $f$ which is
massless~\cite{Hassan:2012wr}. Working in the non-relativistic limit
$c\to\infty$, we find that the massless combination reduces to an
ordinary Poisson equation for the Newtonian potentials $U_g$ and $U_f$
associated with the two metrics, namely
\begin{equation}\label{scalareq}
\Delta\left(\frac{2M_g^2}{\alpha}U_g - \frac{2M_f^2}{\beta}U_f\right)
= -\frac{1}{\alpha}\bigl(\rho^*_\text{bar} +
\rho^*_g\bigr)+\frac{1}{\beta} \rho^*_f\,,
\end{equation}
where $\rho^*_\text{bar}$, $\rho^*_g$ and $\rho^*_f$ denote the
ordinary Newtonian densities of the matter fluids. However, with
massive (bi-)gravity the two sectors associated with the metrics $g$
and $f$ do not evolve independently but are linked together by a
constraint equation which comes from the Bianchi identities. We find
that this constraint reduces in the non-relativistic limit to
\begin{equation}\label{constrNR}
\bm{\nabla}\bigl(\alpha U_g + \beta U_f\bigr) = 0\,.
\end{equation}
Combining~\eqref{scalareq}--\eqref{constrNR} and using~\eqref{Nlimit}
we readily obtain the following Poisson equation for the Newtonian
potential in the ordinary sector,
\begin{equation}\label{poissonNR}
\Delta U_g = - 4\pi \Bigl(\rho^*_\text{bar} + \rho^*_g -
  \frac{\alpha}{\beta} \rho^*_f\Bigr) \,.
\end{equation}
Similarly we find that the equation governing the vector field,
obtained by varying the action~\eqref{lagrangian} with respect to
$\mathcal{A}_\mu$, reduces in the non-relativistic limit to a single
Coulomb type equation,
\begin{equation}\label{coulomb}
\bm{\nabla}\cdot\Bigl[\mathcal{W}_{X}\bm{\nabla}\phi\Bigr] = 4\pi
\Bigl(\rho^*_g - \frac{\alpha}{\beta} \rho^*_f\Bigr)\,,
\end{equation}
where $\mathcal{W}_{X}$ is the derivative of $\mathcal{W}$ with
respect to its argument $X$. Finally we have also at our disposal the
equations of motion of the baryons and the dark matter particles. We
look for explicit solutions of the equations of motion of dark matter
in the form of plasma-like oscillations around some equilibrium
configuration. Thanks to the constraint equation~\eqref{constrNR} we
find that an equilibrium is possible, i.e. the dark matter particles
are at rest, when the Coulomb force annihilates the gravitational
force, namely
\begin{equation}\label{equilibrium}
\bm{\nabla}U_g + \bm{\nabla}\phi=0\,.
\end{equation}
At equilibrium we shall grasp a mechanism of gravitational
polarization, i.e. we can define a polarization field which will be
aligned with the gravitational field. Away from equilibrium we find
that the polarization field undergoes stable plasma like oscillations
(see~\cite{BH15b} for details). Finally, combining the three
equations~\eqref{poissonNR},~\eqref{coulomb} and~\eqref{equilibrium}
we easily deduce that the equation for the ordinary potential $U_g$
takes exactly the Bekenstein-Milgrom form~\cite{BekM84},
\begin{equation}\label{poissonpol}
\bm{\nabla}\cdot\Bigl[\bigl(1-\mathcal{W}_{X}\bigr)\bm{\nabla}
  U_g\Bigr] = - 4 \pi \rho^*_\text{bar} \,,
\end{equation}
with MOND interpolating function $\mu=1-\mathcal{W}_{X}$. Furthermore,
thanks to the postulated form~\eqref{W0} of the function
$\mathcal{W}(X)$ we recover the correct deep MOND regime when $X\to
0$. We refer to~\cite{BBwag,BB14,BH15b} for more details about this
way of recovering the MOND phenomenology, which is of course
reminiscent of the dielectric analogy of MOND.

Let us emphasize that gravitational polarization \& MOND appear as
natural consequences of this model, and are made possible by the
constraint equation~\eqref{constrNR} linking together the two metrics
of massive bigravity theory. Furthermore this requires that the
gravitational force can be annihilated by some internal
non-gravitational force, here chosen to be a vector field. This
implies the existence of a coupling between the two species of dark
matter particles living in the $g$ and $f$ sectors. Unfortunately, as
has been shown in~\cite{BH15b}, the latter coupling is problematic as
it yields a ghostly degree of freedom in the dark matter
sector. Further work is needed to determine the exact mass of the
ghost and see whether the required polarization mechanism and the
ghost absence are compatible. On the phenomenology side, the
cosmological implications of the model and the post-Newtonian
parameters in the Solar System should be investigated in great
details, as it has been done for the previous model~\cite{BB14}.

\section*{References}

\bibliography{DDM_LB_LH.bib}

\end{document}